\documentclass[twoside]{article}
\usepackage{fleqn,espcrc2}

\usepackage{graphicx}
\usepackage{psfig}

\newcommand{\AmS}{{\protect\the\textfont2
  A\kern-.1667em\lower.5ex\hbox{M}\kern-.125emS}}

\title{Vortex lattice structures and pairing symmetry in Sr$_2$RuO$_4$}

\author{D.F. Agterberg,\address{
NHMFL, Florida State University, Tallahassee, Florida 32310, USA}
R. Heeb,\address{Theoretische Physik, ETH Honggerberg, 
CH-8093 Zurich, Switzerland} 
P.G. Kealey, T.M. Riseman, E.M. Forgan, A.P. Mackenzie, L.M. Galvin, R.S. Perry, 
\address{School of Physics and Astronomy, University of Birmingham, Birmingham
B15 2TT, U.K.}
S.L. Lee, \address{School of Physics and Astronomy, 
University of St. Andrews, St. Andrews, Fife,  KY16 9SS, U.K.}
D. McK. Paul, \address{Department of Physics, University of Warwick, 
Coventry CV4 7AL, U.K.}
R. Cubitt, \address{Institut Laue-Langevin, F-38042, Grenoble, France}
Z.Q. Mao, S. Akima, and Y. Maeno \address{Department of Physics, Kyoto
University, Kyoto 606-8052, Japan}}
\begin{document}

\begin{abstract}
Recent experimental results indicate that superconductivity in Sr$_2$RuO$_4$ 
is described by the  $p$-wave $E_u$ representation 
of the $D_{4h}$ point group. Results on the vortex 
lattice structures for 
this representation are presented. The theoretical results are compared 
with experiment.

\vspace{1pc}
\end{abstract}

\maketitle

\section{Introduction}

The oxide Sr$_2$RuO$_4$ has a structure similar to high $T_c$ materials
and was observed to be superconducting by
Maeno {\it et al.} in 1994 \cite{mae94}.
It has been established that this superconductor is not a conventional
$s$-wave superconductor:
NQR measurements
show no indication of a Hebel-Slichter peak in $1/T_1T$ \cite{ish97},
and $T_c$ is strongly suppressed below the maximum value of 1.5 K 
by non-magnetic impurities \cite{mac98}.
More recent experiments indicate 
an odd parity  gap function of the
form ${\bf d}({\bf k})=\hat{z}(\eta_x k_x+\eta_y k_y)$. 
The Knight shift measurements of Ishida {\it et al.} \cite{ish98}
reveal that the spin susceptibility is unchanged upon entering the  
superconducting state; this is consistent with $p$-wave superconductivity 
(as predicted by Rice and Sigrist \cite{ric95}).
Furthermore, these measurements were conducted with the applied field
in the basal plane.   
Since the orientation of the gap function (that is ${\bf d}$)
is orthogonal
to the spin projection of the Cooper pair \cite{sig91}, these measurements are 
consistent with
the gap function aligned along the $\hat{z}$ direction.   
The $\mu$SR experiments of Luke {\it et. al.}
have revealed spontaneous fields in the Meissner state \cite{luk98}. This indicates 
that the superconducting
order parameter must have more than one component \cite{sig91,luk98}.
Naively, the observation of spontaneous fields leads 
to the conclusion that the superconducting gap function  breaks time reversal
symmetry (${\cal T}$) and therefore must
have the form ${\bf d}({\bf k})=\hat{z}(k_x\pm i k_y)$. However,
the muons probe inhomogeneities in the superconducting
state since any bulk magnetic fields are 
screened in the Meissner state. Consequently, ${\cal T}$ 
need only be broken in the vicinity of the inhomogeneities. 
For example, it has been pointed out
in Ref.~\cite{sig91} that a gap function given by 
$\vec{\eta}=(1,\pm 1)$ can give rise to domain walls
that break time reversal symmetry.  In principle this can also lead
to a $\mu$SR  signal as seen by Luke {\it et al.}. 
Consequently, these two experiments 
lead to the gap function ${\bf d}({\bf k})=\hat{z}(\eta_x k_x+\eta_y k_y)$,
with the specific form of the order parameter $\eta_x$ and 
$\eta_y$ undetermined. 
In this talk, the vortex lattice structures arising in 
the phenomenological theory of this $E_u$ representation are 
examined. 
First, the theory is examined for an applied
magnetic field in the basal plane along one of the  two-fold 
symmetry axes. It is shown that
multiple vortex lattice phases are generic to this representation. Then
the weak-coupling limit of the theory is examined for the
applied magnetic field along the $c$-axis. It is shown that a square vortex
lattice results for parameters relevant to Sr$_2$RuO$_4$; consistent 
with the experimental observation of a square vortex lattice \cite{ris98}.
The theory of the square vortex lattice is compared with the observed results
and good agreement is found.
 
\section{Free Energy}

The dimensionless free energy 
for the $E_u$ representation of $D_{4h}$ 
is given by \cite{sig91}
\begin{eqnarray}
f=&-|\vec{\eta}|^2+|\vec{\eta}|^4/2+
\beta_2(\eta_x\eta_y^*-\eta_y\eta_x^*)^2/2 \nonumber \\
& +\beta_3|\eta_x|^2|\eta_y|^2 +
|\tilde{D}_x\eta_x|^2+|\tilde{D}_y\eta_y|^2 \nonumber \\
& +\kappa_2(|\tilde{D}_y\eta_x|^2+
|\tilde{D}_x\eta_y|^2) \label{eq1} \\ &+ \kappa_5(|\tilde{D}_z\eta_x|^2+
|\tilde{D}_z\eta_y|^2)
\nonumber \\ & +\kappa_3[(\tilde{D}_x\eta_x)(\tilde{D}_y\eta_y)^*+h.c.] \nonumber 
\\ & +
\kappa_4[(\tilde{D}_y\eta_x)(\tilde{D}_x\eta_y)^*+h.c.]
+h^2/(8\pi).
\nonumber
\end{eqnarray}
where $\tilde{D}_j=\nabla_j-\frac{2ie}{\hbar c} A_j$,
${\bf h}=\nabla\times {\bf A}$, and ${\bf A}$ is the vector potential.
The stable homogeneous solutions are easily determined. There are three
phases: (a) $\vec{\eta}=(1,i)/\sqrt{2}$ ($\beta_2>0$ and 
$\beta_2>\beta_3/2$), (b) $\vec{\eta}=(1,0)$ 
($\beta_3>0$ and $\beta_2<\beta_3/2$),
and (c) $\vec{\eta}=(1,1)/\sqrt{2}$ ($\beta_3<0$ and $\beta_2<0$). 
Phase (a) is nodeless and phases (b) and (c) have line nodes.
For some of the discussion in this paper the Ginzburg Landau coefficients are
determined within a weak-coupling approximation in the clean limit.
In this case, taking for the $E_{u}$ REP the gap function
described by the pseudo-spin-pairing gap matrix:
$\hat{\Delta}=i[\eta_1 v_x/\sqrt{\langle v_x^2\rangle}+
\eta_2v_y/\sqrt{\langle v_x^2\rangle}]\sigma_z \sigma_y$,
where the brackets $\langle \rangle$ denote an average over the Fermi surface
and $\sigma_i$ are the Pauli matrices, it is  found that \cite{agt98}
$\beta_2=\kappa_2=\kappa_3=\kappa_4=
(\nu+1)/(3-\nu)$ and
$\beta_3=4\nu/(3-\nu)$ where 
\begin{equation}
\nu=\frac{\langle v_x^4\rangle -3 \langle v_x^2v_y^2\rangle}
{\langle v_x^4\rangle + \langle v_x^2v_y^2\rangle}.
\end{equation}
Note that $|\nu|\le 1$ and $\nu=0$ for a spherical or a cylindrical Fermi
surface. Also note that $\vec{\eta}\propto(1,i)$ is the stable homogeneous
state for all $\nu$. 

\section{Magnetic field in the basal plane}

For the $E_u$ model, symmetry arguments imply that 
the vortex lattice phase diagram contains
at least two vortex lattice phases for magnetic fields applied along 
at least two of the four two-fold symmetry axes \cite{agt98}.     
To demonstrate this,  
consider
the magnetic field along the $\hat{x}$ direction
($\hat{x}$ is chosen to be along the crystal ${\bf a}$ axis)
 and a homogeneous zero-field state $\vec{\eta}\propto(1,i)$.
The presence
of a magnetic field along the $\hat{x}$ direction breaks the
degeneracy of the $(\eta_x,\eta_y)$ components, so that only
one of these two components will order
at the upper critical field [{\it e.g.} $\vec{\eta}\propto(0,1)$].
It has been
shown for type II superconductors with a single component
order parameter that the 
solution is independent of $x$ \cite{luk95} so that $\sigma_x$
(a reflection about the $\hat{x}$ direction)
is a symmetry operation of the $\vec{\eta}
\propto (0,1)$ vortex phase.
Now consider the zero-field phase $\vec{\eta}\propto(1,i)$;
$\sigma_x$ transforms $(1,i)$ to $(-1,i)\ne e^{i\psi}(1,i)$
where $\psi$ is phase factor. This implies that $\sigma_x$ is {\it not}
a symmetry operator of the zero-field phase. It follows that there
must exist a second transition in the finite field phase at
which $\eta_x$ becomes non-zero. Similar arguments hold
for the field along any
of the other three two-fold symmetry directions in the basal
plane. Consequently a zero-field state $\vec{\eta}\propto(1,i)$ 
must exhibit two
vortex lattice phases when the field is applied along any of
the four two-fold symmetry axes.   
Similar arguments for  
$\vec{\eta}\propto(1,0)$ or $\vec{\eta}\propto(1,1)$. 
imply that there must exist  
at least two vortex lattice phases for only two of the four two-fold
symmetry axes. For a zero-field state $\vec{\eta}=(1,0)$ fields 
along the $(1,1)$ or the $(1,-1)$ directions 
will result in two vortex lattice phases (for the $(1,0)$ and the $(0,1)$
directions multiple vortex lattice phases may exist, but are not
required by symmetry). For a zero-field state $\vec{\eta}\propto(1,1)$, 
two  vortex lattice phases will exist for fields along
the $(1,0)$ or the $(0,1)$ directions. 

\begin{figure}[t]
\centerline{
\psfig{figure=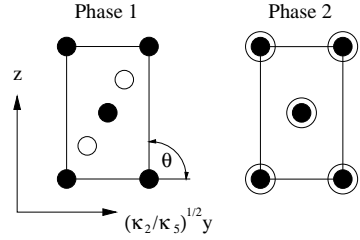,height=1.2in}}
\caption{Stable vortex lattice structures found for the field along $\hat{x}$.
The open (closed) circles correspond to zeroes of the $\eta_x$ ($\eta_y$)
lattice.
\label{fig1}}
\end{figure}

\begin{figure}[t]
\centerline{
\psfig{figure=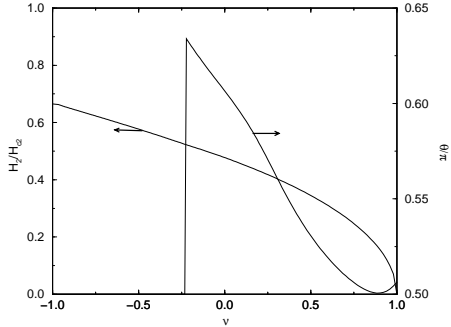,height=2in}}
\caption{The ratio of the two transition fields and $\theta$ 
(the angle relating the orientation of the vortex lattice to the ionic lattice) as
a function of $\nu$.
\label{fig2}}
\end{figure}

An analysis of the free energy of Eq.~\ref{eq1} reveals additional information
about the phase diagram \cite{agt98,kit99}. 
Assuming the large $\kappa$ limit (note $\kappa\approx 30$
for this field orientation in Sr$_2$RuO$_4$), 
the vector potential
can be taken to be ${\bf A}=Hz(\sin\phi,-\cos\phi,0)$  ($\phi$ is
the angle in the basal plane that the applied magnetic field makes with
the $\hat{x}$ direction). The
component of ${\bf D}$ along the field is set to zero.
The upper critical field found by this method exhibits a four-fold
anisotropy in the basal plane \cite{agt98,sig91}.
For concreteness, the field is taken along the $\hat{x}$ direction
($\phi=0$). Introducing the raising and lowering operators 
$\Pi_{\pm}=q(\sqrt{\kappa_2/\kappa_5}D_y \pm i D_z)/\sqrt{2}$ with
$q^2=\sqrt{\kappa_5/\kappa_2}/H$, the gradient portion of the
free energy can be written as
\begin{eqnarray}
f_{grad}=&\sqrt{\kappa_5\kappa_2}H\{\eta_x^*[1+2N]\eta_x +
\eta_y^*[(\frac{1}{2}+\frac{1}{2\kappa_2})\nonumber \\ &
(1+2N)+(\frac{1}{2\kappa_2}-\frac{1}{2})
(\Pi_+^2+\Pi_-^2)]\eta_y\}\nonumber.
\end{eqnarray} 
Assuming that $\kappa_2<1$ the first transition is given by
the standard hexagonal Abrikosov vortex lattice 
solution for $\eta_x$.
To find the second transition the lowest eigenstate 
of $(\frac{1}{2}+\frac{1}{2\kappa_2})
(1+2N)+(\frac{1}{2\kappa_2}-\frac{1}{2})(e^{2i\theta}\Pi_+^2+
e^{-2i\theta}\Pi_-^2)$ must be found (where $\theta$ has
been introduced so that   
the vortex lattice can be rotated  
with respect to the ionic lattice) 
and a vortex lattice solution for $\eta_y$ must be constructed from this.   
Note that the zeroes of the $\eta_x$ lattice and the zeroes
of the $\eta_y$ lattice need not coincide \cite{joy91}. 
This method yields for the  
ratio of the second transition ($H_2$) to the upper 
critical field ($H_{c2}$)
\begin{equation}
\frac{H_2}{H_{c_2}^{ab}}=\frac{\beta_A-\beta_m}
{\beta_A\sqrt{1/\kappa_2}-\beta_m}\label{eq4}
\end{equation}
where $\beta_A=1.1596$, $\beta_m=(1-\beta_2+\beta_3)S_1-|\beta_2S_2|$,
$S_1=\overline{|\eta_1|^2|\eta_2|^2}/
(\overline{|\eta_1|^2}\hphantom{a}\overline{|\eta_2|^2})$,
$S_2=\overline{(\eta_1\eta_2^*)^2}/(\overline{|\eta_1|^2}
\hphantom{a}\overline{|\eta_2|^2})$, the over-bar 
denotes a spatial average,
and $\beta_m$ must be minimized   
with
respect to $\theta$ and the displacement 
between the zeros of the $\eta_x$ and the $\eta_y$
lattices. Diagrams of the predicted vortex structures are shown
in Fig.~\ref{fig1} and the  
phase diagram for the weak-coupling limit is shown in Fig.~\ref{fig2}.  
Two vortex lattice configurations are found to be stable
as a function of $\nu$.
For $-0.23<\nu<1$ (Phase 1), the displacement between the the zeroes of the 
$\eta_x$ and $\eta_y$ lattices is one half of a hexagonal lattice basis 
vector  
and $\theta$ varies with $\nu$ (and in general with field and temperature) 
so that the vortex lattice is not aligned with the ionic lattice.
For $-1<\nu<-0.23$ (phase 2), the $\eta_x$ and $\eta_y$ lattices 
coincide and a vortex lattice basis vector
lies along the $\hat{y}$ direction.
Kita \cite{kit99} has argued that for phase 1, at some field 
below $H_2$ there will be a first order 
transition from phase 1 to phase 2. 
For the field along $\hat{x}\pm\hat{y}$ the phase diagram is given by
replacing $\nu$ with $-\nu$. Recent experiments of Mao {\it et al.} 
\cite{mao00}
show some support for the theory presented here.

\section{Magnetic field along the $c$-axis}

\begin{figure}[t]
\centerline{
\psfig{figure=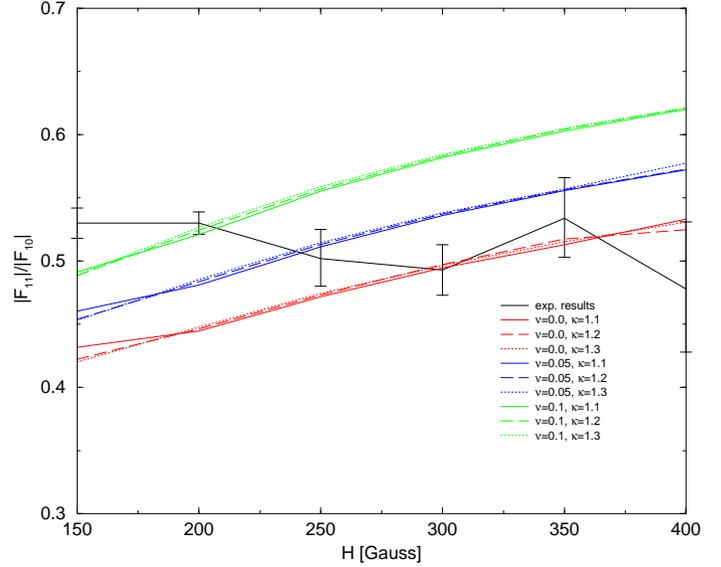,height=3in,angle=270}}
\caption{Comparison between the weak-coupling $E_{u}$ theory 
and experimental results for the ratio of the spatial Fourier 
components of the field corresponding to the $(1,0)$ and  $(1,1)$
Bragg peaks of the square vortex lattice. 
\label{fig3}}
\end{figure}

The free energy in Eq.~\ref{eq1} in the weak coupling limit has been 
used to determine the vortex 
lattice structure for the field along the $c$-axis for fields near the
upper critical field \cite{agt98-2}. The main conclusion of this analysis 
is that vortex lattice is
square for $|\nu|>0.0114$ and that the vortex lattice is oriented along
the crystal lattice for $\nu<-0.0114$ and rotated 45 degrees from the
crystal lattice for $\nu>0.0114$.  The analysis near $H_{c1}$ has also been carried
out with the result that the vortex lattice at $H_{c1}$ will be hexagonal and with
increasing field the lattice will continuously distort until a square vortex 
lattice is formed \cite{hee99}. 
A square vortex lattice has been observed
by Small Angle Neutron scattering (SANS) 
and the orientation implies $\nu>0.0114$ \cite{ris98}. One notable feature of the SANS
measurements is the observation of Bragg peaks beyond the first Bragg peak.   
The analysis of Eq.~\ref{eq1} has 
been extended to fields below $H_{c2}$ and the 
lowest Bragg peaks of the field distribution have been calculated and compared
to experiment. One comparison is shown in Fig.~\ref{fig3} (note that in the SANS
measurements there is flux pinning which is not included in the calculations).  
The large size of the higher order Bragg
peaks relative to that expected for a single complex order parameter 
component (Abrikosov) theory and the agreement between the theory and the
experiment gives support to the $E_u$ theory. However, note that it is
possible that an Abrikosov theory with sufficiently large non-local corrections 
may also account for the observed results.

\section{Acknowledgments}
We acknowledge support from NSF DMR9527035 (DFA), the Swiss Nationalfonds,
the U.K. E.P.S.R.C., and CREST of Japan Science and Technology Corporation.
The neutron scattering was carried out at the Institut Laue-Langevin, 
Grenoble.   
We wish to thank 
T.M. Rice and M. Sigrist for informative discussions.


\begin{thebibliography}{99}
\bibitem{mae94} Y. Maeno {\it et al.}, 
Nature {\bf 372}, 532 (1994).

\bibitem{ish97} K. Ishida {\it et al.}, 
Phys. Rev. B {\bf 56}, 505 (1997).

\bibitem{mac98} A.P. Mackenzie {\it et al.}, 
Phys. Rev. Lett. {\bf 80}, 161 (1998).

\bibitem{ish98} K. Ishida {\it et al.}, 
Nature (London) {\bf 96}, 658 (1998).
 
\bibitem{ric95} T.M. Rice and M. Sigrist, J. Phys.: Condens. Matter {\bf 7},
L643 (1995).

\bibitem{sig91} M. Sigrist and K. Ueda, Rev. Mod. Phys. {\bf 63}, 239 (1991).

\bibitem{luk98} G.M. Luke {\it et al.}, 
, Nature {\bf 394}, 558 (1998).


\bibitem{ris98} T.M. Riseman {\it et al.}, 
Nature {\bf 396}, 242 (1998); and correction in press (2000).

\bibitem{agt98} D.F. Agterberg, Phys. Rev. Lett. {\bf 80}, 5184 (1998).

\bibitem{luk95} I.A. Luk'yanchuk and M.E. Zhitomirsky, Supercond.
Rev. {\bf 1}, 207 (1995).

\bibitem{kit99} T. Kita, Phys. Rev. Lett. {\bf 83}, 1846 (1999).

\bibitem{joy91} R. Joynt, Europhys. Lett. {\bf 16}, 289 (1991);
A. Garg and D.C. Chen, Phys. Rev. B {\bf 49}, 479 (1994).

\bibitem{mao00} Z.Q. Mao {\it et al.}, Phys. Rev. Lett. {\bf 84}, 991 (2000). 

\bibitem{agt98-2} D.F. Agterberg, Phys. Rev. B {\bf 58}, 14 484 (1998).

\bibitem{hee99} R. Heeb and D.F. Agterberg, Phys. Rev. B {\bf 59}, 7076 (1999).  


\end{thebibliography}
\end{document}